\documentstyle[prl,aps,multicol,epsfig]{revtex}

\begin{document}
\large

\title{Efficient Quantum State Tomography for Quantum 
Information Processing using a two-dimensional Fourier 
Transform Technique}

\author{Ranabir Das$^\dagger$, T.S. Mahesh $^\dagger$ and Anil Kumar $^{\dagger,\ddagger}$\\
        $^{\dagger}$ {\it Department of Physics,}
        $^{\ddagger}$ {\it Sophisticated Instruments Facility}\\
        {\it Indian Institute of Science, Bangalore 560012 India}\\}

\maketitle

\begin{abstract}
	A new method of quantum state tomography for 
quantum information processing is described. The 
method based on two-dimensional Fourier transform technique 
involves detection of all the off-diagonal elements of the density matrix 
 in a  two-dimensional experiment. All the diagonal elements are detected in another 
 one-dimensional experiment.  The method is efficient and applicable to a wide range of spin systems. 
The proposed method is explained using a 2 qubit system and demonstrated
 by tomographing arbitrary  complex density matrices of 2 and 4 qubit systems using simulations.
\end{abstract}

\section{Introduction}
Quantum Computation offers exciting possibilities of solving 
complex computational problems using algorithms which exploit
the quantum nature of the system. The idea, first proposed by 
Feynman$\cite{rf}$, is being feverishly  pursued by many [2-6].
Several algorithms like Shor's factorization algorithm,
Grover's search algorithm, Deutsch-Jozsa algorithm, quantum Fourier
transform, quantum counting and quantum error-correction codes have 
been developed  and have clearly established the premise [7-24]. 
The  last step in quantum information processing and 
quantum simulations is the measurement of the output quantum state,
known as quantum state tomography.  In the case of ensemble systems this amounts
to measuring the output density matrix. The output state of a quantum algorithm normally corresponds
to some classical information and therefore, it is sufficient to measure all the diagonal elements of the 
density matrix which corresponds to the probabilities of various 
eigenstates.  However, full quantum state tomography is generally 
carried out wherever possible, because of the following reasons;
(i) knowledge of the full output density matrix allows one to find 
out the experimental errors and to calculate the fidelity of the implementation \cite{ic,ai,ci,tel,ap,qft}, and
(ii) if one wishes to monitor the flow of the  implementation
of an algorithm at any intermediate step, then the best option is to 
measure the full interemediate density matrix \cite{ci}. 

 For an n-qubit ensemble system, the size of the Hilbert space increases as 2$^n$ and the number of 
density matrix elements increases as  2$^n\times$2$^n$.  
 Of these there are $M=(2^{n-1})(2^{n-1}+1)$ independent elements, of which 
 $(n2^{n-1})$ elements are one qubit single quantum observable coherences. To measure the remaining elements,
 a series of one-dimensional experiments with readout pulses to rotate the unobservables
into observables, have been used \cite{ai,ci,ap,qft}. 
Here we propose a new method for quantum state tomography based on the two-dimensional
Fourier transform technique, where all the off-diagonal elements of a density matrix, 
both unobservable and observable, are measured in a two dimensional experiment. 
All the diagonal elements are measured in another one-dimensional experiment. 

 It has been pointed out \cite{refjo} that the earlier method of tomography \cite{ai} involving 
a large number of different measurements; while works well for 
small spin systems, becomes "prohibitively complex" for large spin systems. Such large 
systems can be easily tomographed using the proposed method. It may be mentioned here, that 
the proposed method uses a two dimensional experiment and requires several $t_1$ increments.
 In principle the size of two dimensional data is independent of the number of spins (qubits).
 However, the number of quantus increase linearly with the number of spins. To maintain the 
same resolution for large number of spins (qubits), the number of $t_1$ increments may have to be 
increased appropriately (at best linearly). It may be recalled that the same principle 
is applied to two dimensional NMR of biomolecules, where the size of data is independent 
of the size of biomolecules \cite{er}.    
 The proposed method can also be used for tomography in a wider range of spin systems, i.e. 
quadrupolar or strongly coupled systems, and it requires non-selective r.f. pulses which are devoid of errors caused by the selective pulses used by the earlier method \cite{ai}. The method is explained using a 2-qubit 
system and demonstrated on 2 and 4-qubit systems using simulations. 

\section{The method}
  The method  is based on the technique of indirect detection 
of multiple quantum coherences in NMR spectroscopy by two dimensional Fourier 
transform technique \cite{aue}, wherein all the off-diagonal elements are measured 
in a  two-dimensional experiment (pulse sequence $1(A)$). The diagonal 
elements of the density matrix are measured in another one-dimensional 
experiment (pulse sequence $1(B)$).  
The NMR pulse sequences for the two experiments are, 
\begin{eqnarray}
&(A)&~t_1 - (\pi/2)_y - G_z - \alpha_{-y} - detect(t_2), \nonumber \\
&(B)&~G_z - \beta_y - detect(t_2), 
\end{eqnarray}
where $t_1$ and $t_2$ are the variable time periods of system evolution,
$(\pi/2)_y$, $\alpha_{-y}$ and $\beta_y$ are the rf pulses, and $G_z$
is the field-gradient pulse.

  In experiment 1(A), a given density matrix $\sigma(0)$ is allowed to evolve for a time $t_1$, 
at the end of which a $(\pi/2)_y$ pulse transform every element into all other 
elements of the density matrix, including diagonal elements. The $G_z$ pulse 
dephases the off-diagonal elements averaging them to zero, and retains only the diagonal elements.
  A $\alpha_{-y}$ pulse transforms the diagonal elements into all elements of the density matrix including
single qubit single quantum coherences. These single quantum coherences are then detected as a function 
of time variable $t_2$. A series of experiments are performed by systematic increment of the $t_1$ period 
and the collected two-dimensional time domain data set $s(t_1,t_2)$ is double Fourier transformed yielding 
a two-dimensional frequency domain spectrum $S(\Omega_1,\Omega_2)$. $S(\Omega_1,\Omega_2)$ contains along 
$\Omega_2$ all single qubit single quantum coherences and along $\Omega_1$, 
contribution of every off-diagonal elements of the density matrix to these transitions, dispersed and displayed 
by their specific frequency of evolution in the time-domain $t_1$. Cross-sections parallel to $\Omega_1$ 
at one single quantum resonance frequency can be fitted to $\sigma(0)$, yielding all the off-diagonal 
elements in single two-dimensional experiment. The diagonal elements of $\sigma(0)$ do not 
contribute to the spectrum. 

 To obtain the diagonal elements; experiment 1(B) begins by destroying all 
off-diagonal elements of $\sigma(0)$ by a gradient pulse, then using a small angle detection pulse 
to convert difference in diagonal elements into observable one qubit single quantum coherences 
by linear response. The amplitudes of the coherences allow calculation of all the off-diagonal elements.
The above protocol is explained in the following by explicit calculations on a two qubit system.  
 
\section{two qubit system}
   Consider a two qubit system consisting of two spin 1/2 nuclei 
of Larmor frequencies $\omega_1$ and $\omega_2$, coupled by a weak indirect
coupling J.  The Hamiltonian for the system is,
${\mathcal H}=\omega_1 I_{1z}+\omega_2 I_{2z}+ J I_{1z}I_{2z},$
where $I_{jz}$ $(j=1,2)$ are the spin operators.
A selective rf pulse of angle  $\theta$ and phase $\phi$ on-resonance 
on spin $j$ corresponds to a unitary transform,
$U_{\theta,\phi}^{j}=exp\{-i \theta(I_{jx} sin\phi + I_{jy} cos\phi)\}.$
All quantum algorithms are implemented in NMR by a specific pulse sequence
involving the evolutions under system Hamiltonian and rf pulses.
A general Hermitian complex trace-less deviation density matrix for 2 qubits, 
has 15 independent elements, is spanned by 15 product operators \cite{er}, and can be expressed as; 
\begin{equation}
\sigma(0)= \sum_{k,l}{q_{kl}~I_{1k}I_{2l}},
\label{expprod}
\end{equation}
where $k$ and $l$ can take values 0,1,2,3 corresponding to
$o,x,y,z$ respectively, but not simultaneously 0, and
$I_{1o}=I_{2o}=I$ is unit matrix, and $q_{kl}$ are real coefficients.
All the elements of the density matrix $\sigma(0)$ can be classified
into two groups: (i)  diagonal elements
involving the deviation populations.
 $\Delta P_k$ (deviations from an average population) of various eigenstates 
and (ii) off-diagonal elements involving one-qubit
coherences (1Q elements) and multi-qubit coherences
(zero and double quantum elements for a 2 qubit system). 
\begin{equation}
\sigma (0)=
\left(
\begin{array}{c|c|c|c}
\Delta P_1 & 1Q_1 & 1Q_2 & 2Q_2 \\ \hline
1Q_1^* & \Delta P_2    & 2Q_0 & 1Q_3 \\ \hline
1Q_2^* & 2Q_0^* & \Delta P_3    & 1Q_4 \\ \hline
2Q_2^*   & 1Q_3^* & 1Q_4^* & \Delta P_4 
\end{array}
\right).
\end{equation}.
The trace condition $\sum_k\Delta P_k$=0 yields 15 independent (12  
 off-diagonal and 3 diagonal) elements in this case.
Each 1Q element, known as a single quantum element, corresponds to product operators of 
the type $I_{1x}$, $I_{1y}$, $I_{1x}I_{2z}$, $I_{1y}I_{2z}$, and 
$I_{2x}$, $I_{2y}$, $I_{2x}I_{1z}$, and $I_{2y}I_{1z}$.  Each 2Q element, known as zero or double
quantum element depending on the frequency of evolution 
$\omega_1-\omega_2$ or $\omega_1+\omega_2$, corresponds to an 
expansion in terms of $I_{1x}I_{2x}$, $I_{1x}I_{2y}$, $I_{1y}I_{2x}$ and $I_{1y}I_{2y}$.

\subsection{Measurement of off-diagonal elements}
Effect of the pulse sequence $1(A)$ on any initial
density matrix $\sigma(0)$ can be described as \cite{er},
\begin{eqnarray}
\sigma(0) 
&\stackrel{t_1}{\longrightarrow}& \sigma_1(t_1) = e^{-i{\mathcal H}t_1} \sigma(0) e^{i{\mathcal H}t_1} \nonumber \\ 
&\stackrel{(\frac{\pi}{2})_y}{\longrightarrow}& \sigma_2(t_1) 
= e^{-i(I_{1y}+I_{2y})\pi/2} \sigma_1(t_1) e^{i(I_{1y}+I_{2y})\pi/2} \nonumber\\
&\stackrel{G_z}{\longrightarrow}& \sigma_3(t_1)= {\mathcal P} \sigma_2(t_1)   \nonumber \\
&\stackrel{\alpha_{-y}}{\longrightarrow}& \sigma_4(t_1) = 
e^{i(I_{1y}+I_{2y})\alpha} \sigma_3(t_1) e^{-i(I_{1y}+I_{2y})\alpha} \nonumber \\
&\stackrel{t_2(measure)}{\longrightarrow}& \sigma_5(t_1,t_2)=e^{-i{\mathcal H}t_2} \sigma_4(t_1) e^{i{\mathcal H}t_2}.
\end{eqnarray}
 Here the operator $\mathcal{P}$ projects and retains only the diagonal part of $\sigma_2(t_1)$.
The complex time domain signal $s(t_1,t_2)$ obtained on measurement as a function of $t_2$, is 
$s(t_1,t_2)=Trace[(I_1^++I_2^+)\cdot\sigma_5(t_1,t_2)]$,
which after double Fourier transform gives two-dimensional (2D) spectrum 
$S(\Omega_1,\Omega_2)$ which is a function of the two frequency variables
$\Omega_1$ and $\Omega_2$.  The $\Omega_2$ axis of this spectrum has only
the single quantum (1Q) elements (four transitions in the 2-qubit system).
Along the $\Omega_1$ axis of the 2D spectrum all the off-diagonal elements
of $\sigma(0)$ yield peaks corresponding to their amplitudes ($q_{kl}$) and individual evolution 
frequencies during the time period $t_1$.

Cross-sections of the signal after Fourier Transform with respect to $t_2$,
 but before $t_1$, $[S(t_1,\Omega_1)]$ taken parallel to $t_1$ at $\Omega_2=\omega_1^+$
and $\Omega_2=\omega_2^+$ respectively correspond to,
\begin{eqnarray}
\sigma_4(\omega^+_1)=e^{-t_1/T_2}sin\alpha[ && (q_{10}+q_{13})cos(\omega_1^+t_1) + 
(q_{10}-q_{13})cos(\omega_1^-t_1) \nonumber \\
			 &&  -(q_{20}+q_{23})sin(\omega_1^+t_1) - (q_{20}-q_{23})sin(\omega_1^-t_1) \nonumber \\
			 &&+\frac{1}{2} 
cos\alpha\{ (q_{11}-q_{22})cos(\omega^Dt_1) + (q_{11}+q_{22})cos(\omega^Zt_1) \nonumber \\
			 &&  (q_{12}+q_{21})sin(\omega^Dt_1) - (q_{12}-q_{21})sin(\omega^Zt_1) \}]/4 \nonumber \\
\sigma_4(\omega^+_2)=e^{-t_1/T_2}sin\alpha[ && (q_{01}+q_{31})cos(\omega_2^+t_1) + 
(q_{01}-q_{31})cos(\omega_2^-t_1) \nonumber \\
                         &&  -(q_{02}+q_{32})sin(\omega_2^+t_1) - (q_{02}-q_{32})sin(\omega_2^-t_1) \nonumber \\
                         &&+\frac{1}{2}
 cos\alpha\{ (q_{11}-q_{22})cos(\omega^Dt_1) + (q_{11}+q_{22})cos(\omega^Zt_1) \nonumber \\
                         &&  (q_{12}+q_{21})sin(\omega^Dt_1) - (q_{12}-q_{21})sin(\omega^Zt_1) \}]/4,
\label{sig4exp}
\end{eqnarray}
where $\omega_j^ \pm  = \omega_j \pm J/2$, $\omega^D=\omega_1+\omega_2$,
$\omega^Z=\omega_1-\omega_2$,
 $q_{kl}$ are the coefficients of expansion of $\sigma(0)$ as in
Eq. (\ref{expprod}), and $T_2$ is the transverse relaxation time of various coherences. $T_2$ 
can be different for each coherence, but taken identical here for simplicity. 
On $t_1$ Fourier transformation of expression \ref{sig4exp}, one obtains   
 the two-dimensional (2D) spectrum $S(\Omega_1,\Omega_2)$ in which the cosine 
terms give absorptive and the sine terms dispersive Lorenztian lines. All 
 the coefficients of off-diagonal elements of $\sigma(0)$ 
can be obtained by fitting the cross-sections 
 from the two-dimensional spectrum $S(\Omega_1,\Omega_2)$ (taken parallel to $\Omega_1$) to
the absorptive/dispersive Lorentzians obtained from 
expressions similar to Eq.~(5). Two cross-sections, one at each qubit, are sufficient
to calculate all the off-diagonal elements of the 2-qubit 
density matrix $\sigma(0)$ of Eq. (\ref{expprod}).
For an n-qubit system, there are $2^{n-1}$ cross-sections 
per qubit.  While only one cross-section per qubit is required to map
all the off-diagonal elements, the remaining cross-sections can be used to minimise the errors.  
It may also be noted that the diagonal elements do not interfere with 
the two-dimensional spectrum obtained by pulse sequence 1(A). Similarly  
the undesired `axial peaks' (having zero frequency during $t_1$ period)
arising due to the longitudinal relaxations during 
 $t_1$ period \cite{er} are also suppressed by the present scheme.

 In the two-qubit case, the one-qubit and two-qubit
coherences have different conversion
ratios, namely, $sin\alpha$ and $\frac{1}{4} sin2\alpha$.
As a compromise $\alpha = \pi/4$ has been used here. For qubit systems 
having higher-qubit coherences, appropriate values of $\alpha$ should be 
used which optimizes the intensities of various orders \cite{nmur}.

\subsection{Measurement of diagonal elements}
   The transformations of the density matrix $\sigma(0)$ by the
pulse sequence $1(B)$ are as follows. 
\begin{eqnarray}
\sigma(0) 
& \stackrel{G_z}{\longrightarrow} & q_{30}I_{1z}+q_{03}I_{2z}+q_{33}I_{1z}I_{2z} \nonumber \\
& \stackrel{\beta_y}{\longrightarrow} & q_{30}(I_{1z}cos\beta+I_{1x}sin\beta)
				       +q_{03}(I_{2z}cos\beta+I_{2x}sin\beta) \nonumber \\
				      &&  + q_{33}(I_{1z}cos\beta+I_{1x}sin\beta)(I_{2z}cos\beta+I_{2x}sin\beta) \nonumber \\
& \stackrel{t_2}{\longrightarrow}& Measure.
\end{eqnarray}
 To measure the coefficients under linear response, $\beta$ should be small \cite{er}. 
The measured one-qubit coherences then are 
\begin{equation}
\beta[q_{30}I_{1x} + q_{03}I_{2x} + q_{33}(I_{1x}I_{2z}+I_{1z}I_{2x})],
\label{diagel}
\end{equation}
which can be rearranged as,
\begin{eqnarray}
\beta[
&& (q_{30}+q_{33}/2)(I_{1x}+2I_{1x}I_{2z})
+(q_{30}-q_{33}/2)(I_{1x}-2I_{1x}I_{2z}) \nonumber \\
&&+(q_{03}+q_{33}/2)(I_{2x}+2I_{1z}I_{2x})
+(q_{03}-q_{33}/2)(I_{2x}-2I_{1z}I_{2x})]/2.
\end{eqnarray}
The coefficients of the four terms in the above expression 
are proportional to the intensities of the corresponding four transitions 
of a two qubit system.  After calculating $q_{30}$, $q_{03}$, and $q_{33}$, all the
diagonal elements of the density matrix $\sigma(0)$ can be calculated,
since the diagonal part is equal to $q_{30}I_{1z}+q_{03}I_{2z}+q_{33}I_{1z}I_{2z}$.

  It should be noted that the gradient pulse $G_z$ 
used in pulse sequences 1(A) to 
destroy off-diagonal elements, does not destroy 
homonuclear zero-quantum coherences. In such cases, an extra small delay $\tau _m$ along with $G_z$, 
 randomly varied between each $t_1$ experiment can suppress the homonuclear zero quantum coherence 
\cite{er}. 
 In experiment 1(B), signal averaging using a few randomly varied $\tau_m$ along with 
$G_z$ would suppress the homonuclear zero quantum coherences.

   The above schemes 1(A) and (B) assume ideal r.f. pulses. To correct for errors due to imperfection of the
 r.f. pulses a third one dimensional experiment can be performed to measure the one qubit coherences directly
without application of any pulses. These coherences
can then be used to normalize all other elements of the density matrix measured by
experiments 1(A) and (B).
\section{Simulation}
 To demonstrate the protocol we tomograph an arbitrary complex density matrix with 
simulations. In a 2-qubit system, such a  density matrix is of the form;
\begin{eqnarray}
\sigma(0)= I_{1z}+2.3 I_{2z}+6.7 I_{1z}I_{2z}+I_{1x}+10 I_{1x}I_{2z}+5 I_{1y}+3.5 I_{1y}I_{2z}
          +2.5 I_{1y}I_{2y}  \nonumber \\
          +7.2 I_{1y}I_{1x}+13 I_{1x}I_{2x}+ 1.45 I_{1x}I_{2y}+ 
2I_{2x}+3.45 I_{1z}I_{2x}+6.9I_{2y}+6.753 I_{1z}I_{2y}  \nonumber 
\end{eqnarray}
\begin{eqnarray}
         =\pmatrix{3.325 & 1.8625-5.1383i & 3-3.75i & 2.65-2.1625i \cr
                   1.8625+5.1383i & -2.325 & 3.875-1.4375i & -2-1.625i \cr
                   3+3.375i & 3.875+1.4375i & -1.025 & 0.1375-1.7618i \cr
                   2.65+2.1625i & -2+1.625i & 0.1375+1.7618i & 0.025 }. 
\end{eqnarray}

 We assume the Larmor frequencies of the two qubits (spins) as $\omega_1$=1200 Hz, $\omega_2$=1800 Hz and 
the indirect coupling constant between the  qubits as J=200 Hz.
Experiment 1(A) is performed to obtain all the off-diagonal elements. $\alpha$ was chosen as 45$^o$. 512
t$_1$ increments were performed yielding the 2D spectrum shown in Fig. 1(a). Cross sections
parallel to $\Omega_1$ taken at one of the transitions of
 each qubit are  shown in Fig 1(b) and (c). These cross sections
were fitted to get all the complex off-diagonal elements of the density matrix. 
$T_2$ was taken as 10ms for all coherences. The diagonal elements were mapped
using experiment 1(B) with $\beta = 10^o$, Fig 1(d). 
The real and imaginary parts of the tomographed density matrix are shown respectively in Figs. 1(e) and 1(f). 
The calculated density matrix matches the input density matrix better than 0.01$\%$ for all 
complex elements.
We have also carried out the simulations on a 4-qubit system (Fig. 2) and tomographed 
 the density matrix with  99.7$\%$ fidelity.

\section{Conclusion}
   Two-dimensional nuclear magnetic resonance spectroscopy provides an
 efficient method for the quantum state tomography.  Only an one-dimensional
 and a two-dimensional experiment are required for measuring
all the elements of the density matrix. Since the earlier method requires a series of 
one dimensional experiments with different readout pulses, for large spin systems the approach 
becomes  enormously complex \cite{jo}. However such systems can easily be tomographed using the proposed 
method by aptly increasing the $t_1$ increments. The earlier method uses spin-selective r.f. pulses, 
which requires long-duration irradiation of a paricular spin. During such a  pulse, the unperturbed spins 
 evolve under the Zeeman and coupling interactions, introducing errors due to measurement \cite{djjo,jo}. 
The method described here requires non-selective short-duration r.f. pulses which do not introduce such errors. 
 Search of more qubits has led researchers to use strongly coupled spin-1/2 nuclei and quadrupolar 
nuclei (spin$>$1/2) oriented in liquid crystalline matrices \cite{fun,mulf,nee,mur,mahesh}. 
For such systems the notion of spin-selectivity 
does not apply, but the proposed method based on non-selective pulses can be used for tomography. 
Recently, the method was used to tomograph the states while 
 quantum informtion processing  in weakly and strongly coupled spin systems \cite{mahesh,grov}.  
 This method can also be extended to a 3-dimensional experiment in which 
quantas of various orders are displaced in different 
planes of the 3-D experiment, increasing the detectibility and the resolution of the 
spectrum \cite{rt}.

\section{Acknowledgments}
 Useful discussions with Prof. K.V.Ramanathan and Mr.Neeraj Sinha of our group 
are gratefully acknowledged.

* Author to whom correspondence should be addressed.
e-mail: $\it{anilnmr@physics.iisc.ernet.in}$
\pagebreak
\vspace*{-8cm}
\begin{figure}
\epsfig{file=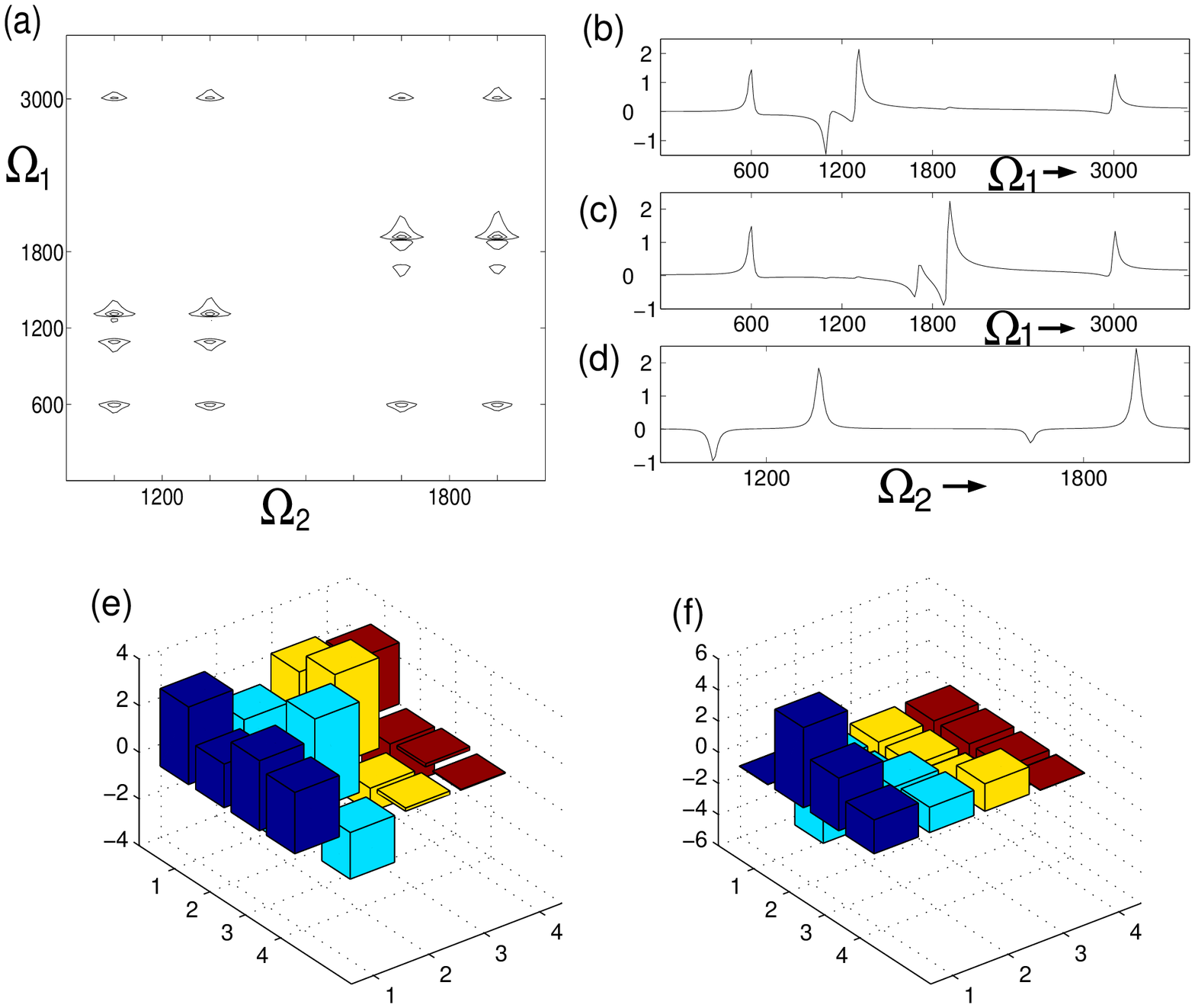,width=20cm}
\caption{Tomography of a complex density matrix in a 2 qubit system. (a) is the 2D spectrum
generated by experiment 1(A) for tomography of all off diagonal elements. (b) is the cross
section parallel to $\Omega_1$ taken at $\Omega_2$= 1300 Hz ($\omega_1+ J/2$) (a transition
frequency of first qubit), (c) is the cross section parallel to $\Omega_1$ 
taken at $\Omega_2$= 1900 Hz ( $\omega_2+J/2$) (a transition
frequency of the second  qubit) and (d) is the 1D spectrum obtained 
by experiment of 1(B) for mapping all diagonal elements of $\sigma(0)$. (e) and (f) are  the real and imaginary part 
of the tomographed density matrix. The calculated values of the tomographed matrix are 
$\sigma_{11}=3.2501$, $\sigma_{12}=1.8625-5.1385i$, $\sigma_{13}=3.0001-3.375i$, $\sigma_{14}=2.6501-2.1624i$, 
$\sigma_{21}=1.8625+5.1385i$, $\sigma_{22}=-2.3251$, $\sigma_{23}=3.8750+1.4374i$, $\sigma_{24}=-2.0001-1.625i$, 
$\sigma_{31}=3.0001+3.375i$, $\sigma_{32}=3.8750-1.4374i$, $\sigma_{33}=-1.025$, $\sigma_{34}=0.1376-1.7618i$, 
$\sigma_{41}=2.6501+2.1624i$, $\sigma_{42}=-2.0001+1.625i$, $\sigma_{43}=0.1376+1.7618i$, and $\sigma_{44}=0.025$. }
\end{figure}

\pagebreak
\begin{figure}
\hspace{-.5cm}
\epsfig{file=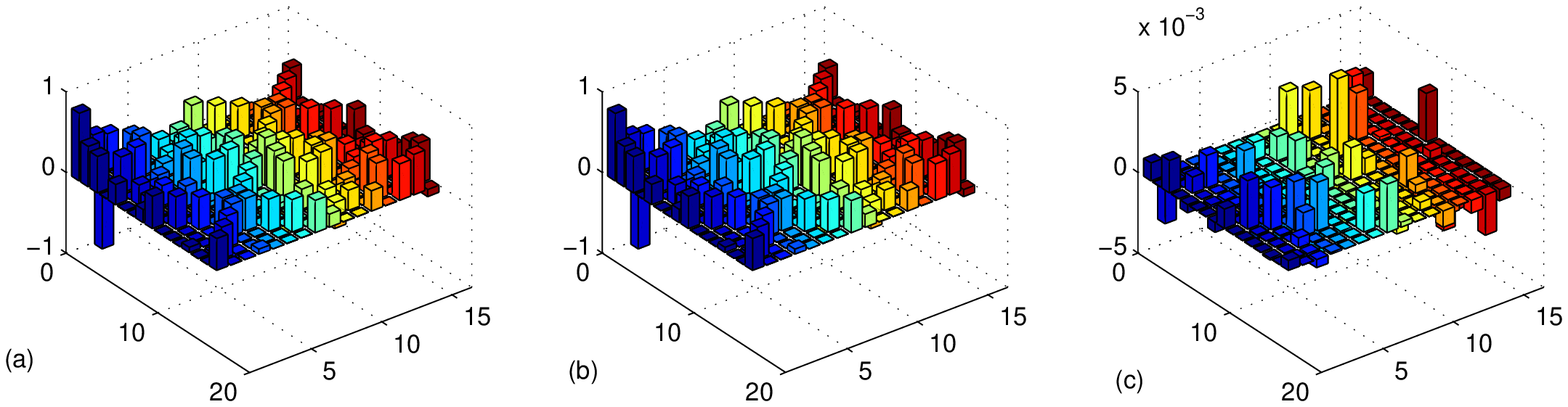,height=18cm,clip=,bbllx=7cm,bburx=15cm,bblly=3cm,bbury=26cm,angle=270}
\end{figure}
\vspace*{-.8cm}
\noindent{\small FIG 2. Tomography of a complex density matrix in 4-qubit system. An arbitrary complex density matrix
$\sigma(0)= 0.8I_{1x}+I_{1y}+0.5I_{2x}+I_{2y}+0.9I_{3x}+1.1I_{3y}+I_{4x}+1.2I_{4y}+
6.3I_{1x}I_{2x}I_{3x}I_{4x}+3.9I_{1x}I_{2y}I_{3y}I_{4y}+ I_{1x}I_{2x}I_{3z}I_{4z}+
1.3I_{2x}I_{3x}I_{4x}$ + $1.9I_{1x}I_{2x}I_{3y}
+1.5I_{1x}I_{2z}I_{3y}I_{4x}+0.6I_{4z}+I_{2z}I_{4z}+1.3I_{2z}I_{3z}I_{4z}+2I_{1z}I_{2z}I_{3z}I_{4z}$,
is reconstructed with the 2D Fourier Transform technique. The frequencies and couplings used in the
simulation are
$\omega_1=600Hz, \omega_2=750Hz, \omega_3=1000Hz, \omega_4=1400Hz, J_{12}=20Hz, J_{13}=10Hz,  J_{14}=70Hz,
 J_{23}=35Hz,  J_{24}=24Hz$ and  $J_{34}=16Hz$. (a) Shows the real part of $\sigma(0)$, (b) is the real
part of reconstructed density matrix and (c) the differences ( magnified by $10^3$) 
between the elements of (a) and (b).
 The 256 complex elements of $\sigma(0)$ (imaginary part not shown here) 
were tomographed with more than 99.7$\%$ accuracy.}

\end{document}